# Case Studies on Superconducting Magnets for Particle Accelerators

*P. Ferracin*[1]
CERN, Geneva, Switzerland

**Abstract**
During the CERN Accelerator School 'Superconductivity for accelerators', the students were divided into 18 groups, and 6 different exercises (case studies), involving the design and analysis of superconducting magnets and RF cavities, were assigned. The problems covered a broad spectrum of topics, from properties of superconducting materials to operation conditions and general dimensions of components. The work carried out by the students turned out to be an extremely useful opportunity to review the material explained during the lectures, to become familiar with the orders of magnitude of the key parameters, and to understand and compare different design options. We provide in this paper a summary of the activities related to the case studies on superconducting magnets and present the main outcomes.

## 1   Introduction

The CERN Accelerator School 'Superconductivity for accelerators', held in Erice from April 24 to May 4 2013, had among its objectives to provide an overview of superconducting RF systems and superconducting magnets for accelerators, to explain the fundamental properties of superconducting materials, and to cover the basic physical principles behind their behaviour, as well as their design and fabrication. In parallel to the series of lectures, a set of cases studies was assigned to the participants, with the goals of complementing with practical work the theories described in the classes and applying analytical formulas, scaling laws, and plots presented in the lectures to solve problems of design and analysis.

A total of six exercises were assigned: four on superconducting magnets and two on RF superconducting cavities. For each of the two areas, a series of subtopics was identified: in the case of superconducting magnets, the work dealt with superconducting strands and cables, magnetic design, operational margins, and mechanical design. In the case of RF cavities, thin films, local defects, and tests of the properties were investigated.

In this paper we give a summary of the case studies for the superconducting magnets, firstly providing a description of the problems and of the learning objectives, then reporting on how the activities were organized, and finally discussing the main results and outcomes.

## 2   Case study problems

The four problems for the case studies on superconducting magnets were formulated as follows.

1. *Low-β $Nb_3Sn$ quadrupole magnets for the HL-LHC*

---

[1] paolo.ferracin@cern.ch

The Large Hadron Collider (LHC) will run at 6.5–7 TeV, providing 300 fb$^{-1}$ of integrated luminosity before the end of the decade. In the 12 years of operation following 2021, CERN is planning to upgrade the LHC to obtain ten times more integrated luminosity, i.e. 3000 fb$^{-1}$. Part of the upgrade relies on reducing the beam sizes in the Interaction Points (IPs) by increasing the aperture of the low-β quadrupole magnets. Currently, the LHC interaction regions feature Nb−Ti quadrupole magnets with a 70 mm aperture and a gradient of 200 T/m.

Exercise: Design a Nb$_3$Sn superconducting quadrupole magnet with an aperture of 150 mm, operating at 1.9 K, and aimed at the upgrade of the LHC Interaction Regions.

2. *Low-β Nb-Ti quadrupoles for the HL-LHC*

   Exercise: Design a Nb−Ti superconducting quadrupole magnet with an aperture of 120 mm, operating at 1.9 K, and aimed at the upgrade of the LHC Interaction Regions.

3. *High-field large-aperture magnet for a cable test facility*

   High-field ($B_{bore} > 10$T)) magnets are needed to upgrade existing accelerators in Europe and to prepare for new projects on a longer time-scale. Nb$_3$Sn is currently the right candidate to meet those objectives because of its superconducting properties and its industrial availability. Over the very long term, further upgrades could require dipole magnets with a field of around 20 T: a possible solution is to combine an outer Nb$_3$Sn coil with an inner coil of High Temperature Superconductor (HTS), which both contribute to the field. In addition, a high-field dipole magnet with a large aperture could be used to upgrade the Facility for REception of Superconducting CAbles FRESCA test facility at CERN, with the aim of meeting the strong demands to qualify conductors at higher fields.

   Exercise: Design a superconducting dipole with a 100 mm aperture that is capable of reaching 15 T at 1.9 K (~90% of $I_{ss}$).

4. *11 T Nb$_3$Sn dipole for the LHC collimation upgrade*

   The second phase of the LHC collimation upgrade will enable proton and ion beam operation at nominal and ultimate intensities. To improve the collimation efficiency by a factor of 15–90, additional collimators are foreseen in the room temperature insertions and in the Dispersion Suppression (DS) regions around points 2, 3, and 7. To provide a longitudinal space of about 3.5 m for additional collimators, a solution based on the substitution of a pair of 5.5 m, 11 T dipoles for several 14.3 m, 8.33 T LHC Main Bending dipoles (MB) is being considered.

   Exercise: Design a Nb$_3$Sn superconducting dipole with a 60 mm aperture and an operational field (80% of $I_{ss}$) at 1.9 K of 11 T.

   For each of the four exercises, a set of more specific questions was assigned as follows.

   - Determine the maximum bore field/gradient and coil size (using sector coil scaling laws).
   - Define strands and cable parameters.
     – Strand diameter and number of strands, Cu/SC ratio, pitch angle, cable width, cable mid-thickness and insulation thickness, filling factor $\kappa$.
   - Determine load-line (no iron), 'short sample' conditions, operational conditions (80% of $I_{ss}$), and margins:
     – $j_{sc\_ss}, j_{o\_ss}, I_{ss}, B_{bore\_ss}$ or $G_{ss}, B_{peak\_ss}$;
     – $j_{sc\_op}, j_{o\_op}, I_{op}, B_{bore\_op}$ or $G_{op}, B_{peak\_op}$;
     – margins in $T, j_{sc}, B_{peak}$.

- Compare 'short sample' conditions, 'operational' conditions, and margins if the same $Nb_3Sn$ (Nb–Ti) magnet uses a Nb–Ti ($Nb_3Sn$) superconductor.

- Define a possible coil lay-out to minimize field errors.

- Determine electromagnetic forces $F_x$ and $F_y$ and the accumulated stress on the coil mid-plane in the operational conditions (80% of $I_{ss}$).

- Evaluate dimensions of iron yoke, collars, and shrinking cylinder, assuming that the support structure is designed to reach 90% of $I_{ss}$.

In addition, the students were asked to compare and evaluate different designs and technological options currently under investigation in the superconducting magnet community.

- High-temperature superconductor: YBCO vs. Bi2212.

- Superconducting coil design: block vs. cos-theta.

- Support structures: collar based vs. shell based.

- Assembly procedure: high coil pre-stress vs. low coil pre-stress.

## 3    Learning objectives

The general purpose of the case studies was to guide the students towards the conceptual design of a superconducting magnet, and, more specifically, towards the definition of its key parameters and dimensions by the use of analytical formulas and scaling laws provided during the lectures. Starting with a set of magnet specifications, the exercises were conceived in such a way that a 'first-order' characterization of different magnet components (strand, cable, coil, yoke, support structure) had to be performed, and the physics behind different design options analysed.

Another characteristic of the case studies was that the specifications chosen for each problem were those of magnets currently under development. The idea was for the students to face the design issues that magnet designers working on different projects are dealing with right now, and for them to be aware of the activities being performed by the different magnet groups around the world.

The first exercise referred to MQXF [1], a $Nb_3Sn$ quadrupole magnet being designed by CERN and the US LARP collaboration [2] for the Interaction Regions of the High-Lumi LHC [3]. The specifications of the second problem were based on MQXC [4, 5], a Nb–Ti quadrupole magnet developed by CERN and CEA Saclay for a future upgrade of the LHC low-$\beta$ quadrupoles [6] and currently under test [7]. The 11 T magnet [8, 9], planned for the upgrade of the LHC collimation system and under fabrication and test at FNAL and CERN, was used to define the third problem. The final problem referred to FRESCA2 [10, 11], a large-aperture $Nb_3Sn$ dipole which is being developed by a CERN–CEA Saclay collaboration for the upgrade of the FRESCA cable test facility.

Each of the exercises started with a request to define the overall coil dimensions for the specified field and aperture by using the scaling laws presented in the lecture 'Magnetic design of SC magnets' by E. Todesco. A detailed description of these laws can be found in [12, 13]. Then, the students were supposed to determine the general cable properties by using plots and data that corresponded to typical superconducting cables provided in the lectures 'Superconductors for magnets I-II' by R. Flukiger and 'Superconducting cables' by P. Bruzzone. Once the sizes of the cable and coil were defined, the following step was to compute magnet load-lines and field-current–temperature margins, as well as to compare the limit performance of Nb–Ti vs. $Nb_3Sn$ magnets utilizing the parameterization of the critical curves for Nb–Ti [14] and $Nb_3Sn$ [15] superconductors. The lecture on magnetic design by E. Todesco also contained analytical formulas to determine the angles of the coil blocks and Cu wedges to minimize field errors, to establish the dimensions of the iron yoke, and, as a result, to outline a preliminary coil and magnet lay-out. The electromagnetic forces and coil stresses were computed by using the sector-coil formulas discussed in the lecture 'Mechanical design of SC magnets I-II' by F. Toral, who adopted the analytical approaches presented in [16–21]. The calculation of the forces allows an estimation of the thickness of outer cylinder, thus completing a simplified conceptual design of the magnet.

## 4    Organization of the activities

The students were divided into 18 working groups with 5 to 6 students per group. As there were 6 exercises, the same problem was assigned to 3 groups. The group members were chosen to have different backgrounds and expertise, and they all came from different universities/institutions. Two afternoons were devoted to the case studies, and all the activities took place in a common room. Such a set-up significantly facilitated the interaction among the different groups and between students and teachers.

The schedule of the school was organized so that all the material required for the exercises was presented and explained in lectures that preceded the case study afternoons. This approach resulted in the exclusions from the case study topics of several subjects that were treated in the latter lectures of the school, such as, for example, quench protection, cryogenics, and fabrication issues. On the last day, the students were requested to give a 10 minute presentation summarizing the work to the other groups and lecturers.

## 5    Outcomes and conclusions

The afternoon sessions dedicated to the case studies were characterized by an extremely active participation by all the students, with a very high level of interaction within each group, among groups, and with the lecturers. Indeed, the teachers clarified some critical aspects of the problems and corrected possible errors during the initial set-up of the solutions. In addition, the lecturers assisted in finding the useful formulas for the exercise amongst the large quantity of material provided during the 10 days of classes. All 18 groups completed the assignments after about 8 to 10 hours of work, and they summarized the results in the final presentations.

Although the general cable and coil parameters were easily identified, computation of the magnet limits through the interception of the load-line with the superconductor critical curves required more supervision. It turned out that concepts such as filling factor, overall current density and short-sample current, which are routinely used by magnet designers, are not straightforward for newcomers, and they needed to be defined properly and thoroughly explained during the classes. The students demonstrated particular interest in the representation of the field in a series of harmonics and in the minimization of the field perturbations via the subdivision of the coil in blocks and Cu wedges. Concerning the mechanical design, the computation of the forces in operational condition was extremely useful in pointing out the general level of stress in a superconducting coil (~30–70 MPa for Nb–Ti, 100–150 MPa for $Nb_3Sn$). Finally, the discussion about alternative design and material options currently considered within the superconducting magnet community allowed the students to formulate, and include in their presentations, interesting comparisons between the possible choices, to outline the pros and cons, and, sometimes, to express their preferences.

The final presentations constituted a unique opportunity for the groups to describe the work carried out during the previous days and to underline the most important conclusions of the design analysis. The various talks were well organized, concise, and informative, and they demonstrated the effectiveness of the case studies as a tool to review the material covered during the lectures, pick the key analytical formulas, identify the magnitude of the different parameters, and outline a preliminary magnet lay-out.